# Clustering in anomalous files of independent particles

Ophir Flomenbom

*Flomenbom-BPS, 19 Louis Marshal St., Tel Aviv, Israel 62668*

**ABSTRACT**

The dynamics of classical hard particles in a quasi one-dimensional channel were studied since the 1960s and used for explaining processes in chemistry, physics and biology and in applications. Here we show that in a previously un-described file made of anomalous, independent, particles (with jumping times taken from, $\psi_\alpha(t) \sim t^{-1-\alpha}$, $0 < \alpha < 1$), particles form clusters. At steady state, the percentage of particles in clusters is about, $\sqrt{1-\alpha^3}$, only for anomalous $\alpha$, characterizing the criticality of a phase transition. The asymptotic mean square displacement per particle in the file is about, $\log^2(t)$. We show numerically that this exciting phenomenon of a phase transition is very stable, and relate it with the mysterious phenomenon of rafts in biological membranes, and with regulation of biological channels.

**INTRODUCTION**

File dynamics (sometimes called, single file dynamics) is the diffusion of $N$ ($N \to \infty$) identical Brownian hard spheres in a quasi-one-dimensional channel of length $L$ ($L \to \infty$) [1-19], such that the spheres do not jump one on top of the other, and the average particles' density is about fixed. The most well know statistical property of this process is that the mean square displacement (*MSD*) of a tagged particle in the file follows, $MSD \approx t^{1/2}$. Indeed, file dynamics were used in modeling numerous microscopic processes [20-26]: the diffusion within biological and synthetic pores and porous material [20, 21, 25], the diffusion along 1D objects, such as in biological roads [26], the dynamics of a monomer in a polymer [22], etc. Nevertheless, in real files, one, or several, of the conditions defining the *basic* file may break down. Studies of generalizations of basic files show a rich spectrum of properties. For example, when the particles can bypass each other with a constant probability upon encounter [9], an enhanced diffusion is seen. When the particles interact with the channel, a slower diffusion is observed [16]. For Brownian files with an initial particles' density law that is not fixed, the diffusion is enhanced [10]. Moreover, in heterogeneous files with diffusion coefficients drawn from a density that diverges like a power law around the origin, slower dynamics are almost always obtained [12-14]. (See part A in the supplementary material (SM) that accompanied this paper for further mathematical description on the files introduced in this paragraph and in the next one). Generalizations of the basic file are important since these models represent reality much more accurately than the basic file.

**THE MODEL**

**Anomalous files of independent particles.-** Only recently, files that are anomalous were studied [17-18]; in such files, the jumping times of the particles are taken from a jumping time probability density function (PDF) of the form: $\psi_\alpha(t) \sim t^{-1-\alpha}$, $0 < \alpha < 1$. In [16], it was shown that in renewal-anomalous files, were all the particles attempt a jump together, the *MSD* scales as the *MSD* of the corresponding Brownian file in the power of $\alpha$. Here, we study previously un-described anomalous files made of independent particles. In such files, a random anomalous time is independently assigned for each particle. The fastest particle attempts a jump, and then, all the random times are adjusted. Finally, the particle that attempted jumping receives a new random time. This system has *N* independent anomalous clocks, where a renewal-anomalous file has only one clock. This is the origin for very different dynamical behaviors: Since the clocks are anomalous and independent, the particles are further connected in space, causing further slowness, even relative with renewal-anomalous files. Mathematically, the reason is that at large times, the order of the jumps that enables motion is exponentially small (with the number of particles that are suppose moving). The basic manifestation of this is a logarithmic scaling with the time of the *MSD* per particle, $MSD \sim ln^2(t)$. Moreover, and even more exciting, we find a unique phenomenon in such files: the formation of clusters. We characterize the criticality of this phase transition showing that the number of particles in clusters at steady state follows, $\sqrt{1-\alpha^3}$. We also prove in many numerical tests that this phenomenon is indeed stable. Finally, we also suggest a link of this phenomenon with the mysterious phenomenon of rafts in membranes [24], and with regulation of biological channels [25].

**RESULTS**

**Scaling law for anomalous files of independent particles.-** Here, we study anomalous files of independent particles using scaling laws. Firstly, we write down the scaling law for the mean absolute displacement (*MAD*) in a renewal file with a constant density [10, 14, 18]:

$$<|r|> \sim <|r|>_{free}/n. \qquad (1)$$

Here, $n$ is the number of particles in the covered-length $<|r|>$, and $<|r|>_{free}$ is the MAD of a free anomalous particle, $<|r|>_{free} \sim t^{\alpha/2}$. In Eq. (1), $n$ enters the calculations since all the particles within the distance $<|r|>$ from the tagged one must move in the same direction in order that the tagged particle will reach a distance $<|r|>$ from its initial position. Based on Eq. (1), we write a generalized scaling law for anomalous files of independent particles:

$$<|r|> \sim \frac{<|r|>_{free}}{n} f(n) \quad ; \quad 0 < f(n) < 1. \tag{2}$$

The first term on the right hand side of Eq. (2) appears also in renewal files; yet, the term $f(n)$ is unique. $f(n)$ is the probability that accounts for the fact that for moving $n$ anomalous independent particles in the same direction, when these particles indeed try jumping in the same direction (expressed with the term, $<|r|>_{free}/n$), the particles in the periphery must move first so that the particles in the middle of the file will have the free space for moving, demanding faster jumping times for those in the periphery. $f(n)$ appears since there is not a typical timescale for a jump in anomalous files, and the particles are independent, and so a particular particle can stand still for a very long time, substantially limiting the options of progress for the particles around him, during this time. Clearly, $0 < f(n) < 1$, where $f(n) = 1$ for renewal files since the particles jump together, yet also in files of independent particles with $\alpha > 1$, since in such files there is a typical timescale for a jump, considered the time for a synchronized jump. We calculate $f(n)$ from the number of configurations in which the order of the particles' jumping times enables motion; that is, an order where the faster particles are always located towards the periphery. For $n$ particles, there are $n!$ different configurations, where one configuration is the optimal one; so, $\frac{1}{n!} \leq f(n)$. Yet, although not optimal, propagation is also possible in many other configurations; when $m$ is the number of particles that move, then, $f(n) \sim \binom{n}{m}(n-m)!\frac{1}{n!}$, where $\binom{n}{m}(n-m)!$ counts the number of configurations in which those $m$ particles around the tagged one have the optimal jumping order. Now, even when $m \sim n/2$, $f(n) \sim e^{-n/2}$. Using in eq. (2), $f(n) \sim e^{-n/n_0}$ ($n_0$ a small number larger than 1), we see,

$$MSD \sim \left(\frac{\alpha}{n_0}\right)^2 ln^2(t). \tag{3}$$

(In eq. (3), we use, $MSD \sim MAD^2$.) In fig. 1, we show that results from simulations coincide with eq. (3), for various values of $\alpha$. Equation (3) shows that asymptotically the particles are extremely slow in anomalous files of independent particles.

**Numerical results of anomalous files of independent particles.-** For understanding this slowness even better, we perform extensive numerical simulations. In the simulations, $N = 501$, and the initial density is constant with a distance of unity among the point particles. At the edges, reflecting boundaries are positioned at points, $\pm 253$. (We use units without dimensions all over). Random jumping distances are distributed uniformly in about a unit interval centered on the origin, and the reflection method is used in moving the particles, namely, a jump is made and the particles' order remains. Simulations were performed for seven values of $\alpha$ in the range of anomaly, $0.024 \leq \alpha \leq 0.9$ (in this range, the average of $\psi_\alpha(t)$ is infinite). In addition, we performed two control simulations: one for a file of independent particles with $\alpha = 3.37$ [that has a finite average for $\psi_\alpha(t)$] and one for normal dynamics. Trajectories obtained from simulations are shown in fig. 1 as a function of the number of the cycles t, where a cycle contains $N$ attempts of jumping. The trajectories exhibit the phenomenon of clustering: namely, particles attract each other and then move pretty much together. It is also evident that the value of $\alpha$ and the number of cycles determine the degree of clustering in the system. We note that the results presented here are independent of the value for $N$ and are qualitative identical for files with finite size particles (see part B of the SM that accompanied this paper).

Characterizing the formation of the clusters, fig. 2A shows $p_n(t, \alpha)$: the percentage of particles in a cluster at t for a particular $\alpha$ (namely, the number of particles in clusters above the total number of particles). Here, when adjacent particles are at a distance not larger than 0.1, they are considered clustered. The curves height depends on $\alpha$, yet when normalizing $p_n(t, \alpha)$ with $\xi(\alpha)$ [$\equiv p_n(t \to \infty, \alpha)$], the curves pretty much coincide with each other (fig. 2B). (In action, $\xi(\alpha)$ is the average of the last 10% of the trajectory.) $\xi(\alpha)$ is shown in fig. 2C with the optimal (4-parameter) fitting function, $\tilde{\xi}(\alpha) = 0.98\left(1 - \left(\frac{\alpha}{0.99}\right)^{3.09}\right)^{0.537} - 0.028$. This fitting function is of the form of,

$$\tilde{\xi}(\alpha) \approx \sqrt{1 - \alpha^3}, \tag{4}$$

When $\alpha \to 0$, almost all particles are in clusters. The fluctuations in $\xi(\alpha)$ are about 5% for $\alpha = 0.9$, and are about 0.5% for $\alpha = 0.024$ (with about a linear interpolation with $\alpha$). The fluctuations in $\xi(\alpha)$ represent the motion of particles among clusters. Namely, for a small value of $\alpha$ at steady state, the particles in a cluster move together, where at larger values of $\alpha$, about 5 percents of the particles diffuse among clusters. Since clustering occurs only for anomalous $\alpha$, $\tilde{\xi}(\alpha)$ describes the criticality of a phase transition. Indeed, $\tilde{\xi}(\alpha)$ has a typical form for a scaling function in critical phenomena [27] (see the next paragraph for further discussion about this point). Complementary information about the clustering is obtained from two additional functions: $p_c(t, \alpha)$ and $S_c(t, \alpha)$. Figure 3A presents $p_c(t, \alpha)$: the percentage of clusters (measured in terms of the number of particles) at t for a particular $\alpha$. For relatively large values of $\alpha$, the number of clusters is also large (yet, the clusters are smaller in size). The fluctuations in the number of clusters is also larger when $\alpha$ is larger. This is in accordance with the behavior of $p_n(t, \alpha)$. Panel 3B shows $\pi(\alpha)$ [$\equiv p_c(t \to \infty, \alpha)$] versus $\alpha$ for all anomalous values of $\alpha$. The optimal fitting function has the form, $\tilde{\pi}(\alpha) = 0.78\left(1 + \left(\frac{\alpha}{1.19}\right)^{2.49}\right)^{0.42} - 0.75$. $\tilde{\pi}(\alpha)$ follows closely a function of the form,

$$\tilde{\pi}(\alpha) \approx 0.6\left(\sqrt{1.7 + \alpha^3} - 1.25\right). \tag{5}$$

$\tilde{\xi}(\alpha)$ and $\tilde{\pi}(\alpha)$ have complementary physical interpretation, seen in their scaling laws following (about), $\sqrt{1 \pm \alpha^3}$. $\tilde{\xi}(\alpha)$ quantifies particles in clusters, where the same number of particles can exist for a small or a large number of clusters. $\tilde{\pi}(\alpha)$ simply counts clusters, and can have the same value when these are either small clusters or large clusters. Importantly, when clustering occurs, we see a small number of large clusters as $\alpha$ becomes smaller, where in a system without clustering, we may see about 10% of small clusters. Figure 3C presents the average size of a cluster, $S_c(t, \alpha)$ $[\equiv p_n(t, \alpha)/p_c(t, \alpha)]$. Here, fluctuations are larger when $\alpha$ is small. Panel 3D shows $\chi(\alpha)$, the asymptotic value of the a cluster's size $[\chi(\alpha) = S_c(t \to \infty, \alpha)]$, with its simple fitting function,

$$\tilde{\chi}(\alpha) = (33 - 37\alpha)/N. \tag{6}$$

Interestingly, the average size of a cluster is limited with about 33 particles when $\alpha \to 0$, where clustering disappears when $\alpha \to 1$, further quantifying the phase transition.

**DISCUSSION AND CONCLUSIONS**

**Characterizations of the clustering.-** Firstly, we recall that slowness is expected in files of anomalous independent particles since the order of the jumps that enables motion is exponentially small (with the number of particles that are suppose moving) and the dynamics are without a typical timescale. For further explaining the clustering, we look on the actual values of the jumping times of the particles after a while that the process has been going on; see fig.4. (These are the quantities discussed in the derivation of the *MSD*.) It is clear from fig. 4 that when $\alpha$ decreases the typical value for a jumping time increases (here, the typical time is the jumping time of most of the particles). The interesting issue here is that when $\alpha$ decreases there is a phenomenon that only a few particles are significantly faster relative to all the others. This tells the story of the clustering and the phase transition: when one particle jumps over and over and over again, it clusters the particles around him, since when only a particular particle moves repeatedly several times, it closes the gap among the particles around him, such that they are and eventually clustered.

How can we explain the form of the fitting functions? Firstly, we note that the fitting function of $\xi(\alpha)$ has a standard form for a scaling function at criticality of a phase transition [27]: $f(\alpha) \sim (1-\alpha)^\mu$ (where a function in $\alpha$ can replace $\alpha$ in generalizations), and $\chi(\alpha)$ and $\pi(\alpha)$ pretty much follow from $\xi(\alpha)$. $\pi(\alpha)$ is complementary to $\xi(\alpha)$, since it has such a physical interpretation, and $\chi(\alpha)$ is the ratio of the previous ones.

Now, for further supporting the form of the fitting function of $\xi(\alpha)$, we calculate the PDF of slowest jumping time when there are $n+1$ jumping times in the band:

$$f_{s.t.}(t; n+1) = \psi(t) \left(\int_0^t \psi(s)ds\right)^n \sim t^{-1-\alpha} e^{-nt^{-\alpha}}. \tag{7}$$

We emphasize the following three points: (1) $f_{s.t.}(t; n+1)$ is very small for times smaller than, $t^* \equiv n^{1/\alpha}$, that is the time when the argument of the exponent $e^{-nt^{-\alpha}}$ is unity. (2) $t^*$ is the typical timescale for most of the particles in the file, in the limit of many cycles. This is indeed seen in fig. 4. The reason is very simple: after many cycles, most of the particles are extremely slow, since only the fast ones move and after several jumps the anomalous properties of $f_{s.t.}(t; n+1)$ 'assign' the particle a very slow jumping time. (3) When calculating the first and the second moments of $f_{s.t.}(t; n+1)$ in the range $t \leq t^*$, we find, $<t> \sim n^{1/\alpha-1}$ and $<t^2> \sim n^{1/\alpha-2}$. This should reflect the properties of the fast particles until the time $t^*$. It is evident that a transition occurs in the second moment when $\alpha > 1/2$: $<t^2>$ vanishes when $\alpha > 1/2$, yet scales with $n$ when $\alpha < 1/2$. Namely, for $\alpha < 1/2$ many of the fast particles are slower than $t^*$, yet when $\alpha > 1/2$, most of the fast particles are indeed faster than $t^*$. This behavior is indeed seen in fig. 4: when $\alpha < 1/2$ fewer and fewer particles are seen in the range $t \leq t^*$, yet when $\alpha > 1/2$ we see many particles in this range. This is the origin for many small clusters when $\alpha > 1/2$ and only a few clusters, yet larger, when $\alpha \leq 1/2$. $\tilde{\xi}(\alpha)$ and $\tilde{\pi}(\alpha)$ capture this property.

**Anomalous files, rafts and channels.-** Now, we also find that clustering is seen in anomalous files embedded in two-dimensions, creating a network of isotropic files, like streets and junctions. Indeed, this system is a generalization of a 1-dimensional file, and is defined with two free parameters: the percentage of intersections (without directional preference in intersections) and the length of the interval until an intersection occurs. We study files that intersect each other for 1% every interval of 10 (see part C in the SM that accompanied this paper for a comprehensive analysis). Among other results, we find that in such a system 50% of the particles are in clusters when $\alpha \to 0$. Indeed, the results are sensitive to the branching parameters: when branching occurs in smaller intervals, clustering decreases, and we can speculate that when diffusion happens in two dimensions (not in a network of one-dimensional files), the clustering phenomenon is not observed when the density is reasonable (not too high). This is in accordance with known results showing that the slower diffusion so typical for a particle in a file in one-dimension does not hold for diffusion of hard particles in two-dimensions, where in such a system a standard diffusion is seen (when the density is not too high). Still, we have chosen here reasonable parameters for the branching: the average size of a jump is 0.25, and the branching occurs every interval of 10; this is not too small interval so branching indeed has a role (seen also in the results), still the branching happens after frequent enough jumps and the clustering is indeed seen.

An isotropic network of files embedded in two-dimensions enables relating the clustering with rafts: a raft in a (two-dimensional) membrane is a dense patch of specific lipo-molecules [24]. The mechanism of the formation of these patches is still not clear, yet it is known that rafts do not largely occur due to an electrostatic attraction. We think that the phenomenon of clustering in anomalous files of independent particles can explain rafts in membranes: given that the lipo-molecules diffusion is anomalous (anomalous diffusion is common in membranes), they will form rafts, since diffusion in biological membranes is describable with the model of an isotropic network of files in two dimensions.

Finally, we expect that the clustering phenomenon is universal and holds in a wide range of external conditions, since the diffusion coefficient of the particles does not affect this phenomenon, yet $\alpha$, the only other external parameter here, is the control parameter. Since clustering is expected universal, it may be used in regulating biological channels, an important topic in biophysics, e.g. [25]; this is achieved when controlling the phase of the anomalous particles in the channel (clustered or diffusing), using one of two possibilities: changing $\alpha$ (smaller or larger than 1) or controlling the synchronization of the particles (synchronized or independent with anomalous $\alpha$).

**FIGURE LEGENDS**

**Figure 1** Nine trajectories from anomalous file of independent particles plotted as a function of the cycle index, t; note that the actual time obeys the formula: $t \approx \text{t}^{1/\alpha}$. Particles are initially positioned at the integers, here shown particles located initially in the range, 122-130. In the simulations, $N = 501$, $\Delta = 1$ (the initial distance among particles), $D = 1$ (the diffusion coefficient of the particles), $dt = 0.13$, and the jumping distance obeys, $\sqrt{2Ddt}(2q - 1)$, where $q$ is a random number uniform in the unit interval. (Here, we use units without dimensions.) The upper panels show trajectories for $\alpha = 9/10$ and the lower panels for $\alpha = 5/10$. Left panels show high resolution trajectories at the initial stage of the process. Right panels show trajectories at low resolution at the last third of the simulation (we plot the trajectory every seven thousand cycles). Trajectories in a cluster look in this plot as one trajectory. Clearly, trajectories attract each other stronger at small values of $\alpha$. This is evident at short times and at large times. We also show the *MSD* for two values of $\alpha$ with fitting functions taken from eq. (3).

**Figure 2** $p_n(t, \alpha)$, its normalized form and $\xi(\alpha)$. (**A**) $p_n(t, \alpha)$ as a function of the event index t, for 10 values of anomalous $\alpha$, $\alpha = 0.9, 0.8, 0.7, 0.6, 0.5, 0.4, 0.3, 0.2, 0.1, 0.024$, for the third (from the bottom) and on curves respectively, and the control curves: an anomalous file with $\alpha = 3.37$ and a normal dynamics file (these curves are most lower ones and pretty much coincide with each other). The clustering phenomenon is unique for anomalous files of independent particles, representing a phase-transition depending on $\alpha$. (**B**) Normalizing $p_n(t, \alpha)$ with its asymptotic value $\xi(\alpha)$, all curves follow pretty much the same route. (**C**) $\xi(\alpha)$ is shown on the right with its fitting curve, $\tilde{\xi}(\alpha)$. As $\alpha$ goes to zero, about 97% of the particles are in clusters.

**Figure 3** $p_c(t, \alpha)$, $\pi(\alpha)$, and $S_c(t, \alpha)$ and, $\chi(\alpha)$. (**A**) $p_c(t, \alpha)$ as a function of the event index t, for 4 values of anomalous $\alpha$, $\alpha = 0.7, 0.5, 0.3, 0.024$, counting from the top curve. The percentage of clusters is smaller when $\alpha$ is small, since the clusters are larger at small $\alpha$. (**B**) $\pi(\alpha)$, the steady state value for the number of clusters in percentages is shown with its fitting curve, $\tilde{\pi}(\alpha)$. As $\alpha \to 0$, the percentage of clusters is 3%. (**C**) $S_c(t, \alpha)$ as a function of the event index t, for the 4 values of anomalous $\alpha$ in (**A**), counting from the lower curve. The average size of a cluster is large when $\alpha$ is large. Here, the average cluster can contain, momentarily, about 10% of the particles. (**D**) $\chi(\alpha)$, the steady state value for the average size of a cluster (in percentage) with its simple fitting curve, $\tilde{\chi}(\alpha)$. As $\alpha \to 0$, the average cluster's size is 33.

**Figure 4** The logarithm of the band of jumping times after about 7 hundred thousand cycles for a system of about 500 point particles for several values of $\alpha$.

# FIGURES

## FIGURE 1

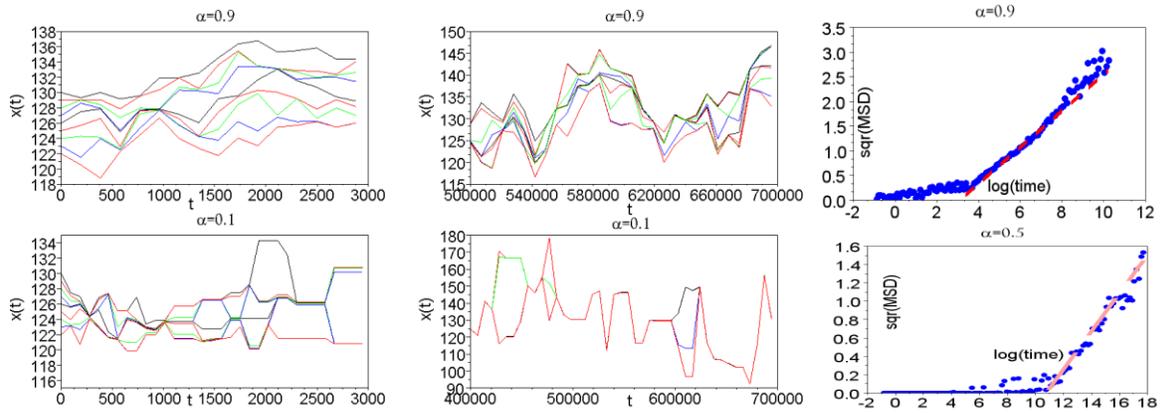

**FIGURE 2**

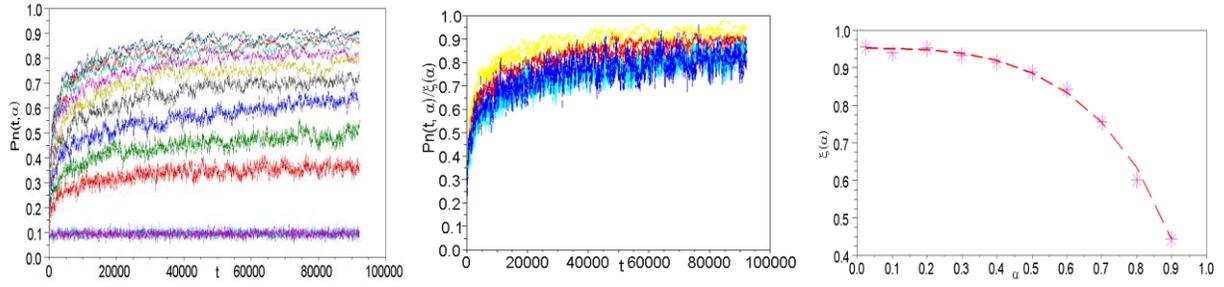

**FIGURE 3**

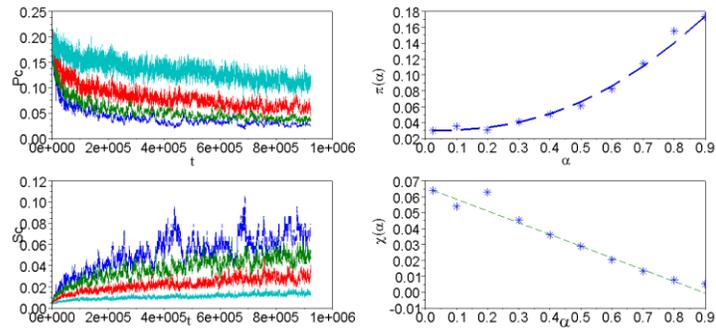

**FIGURE 4**

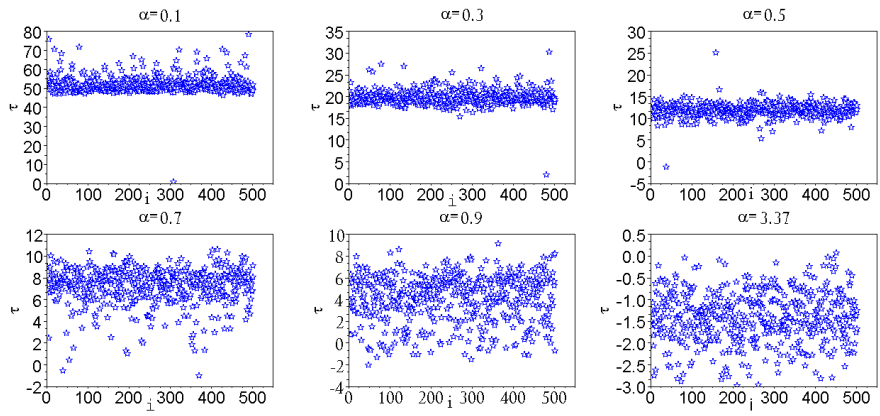

# Supplementary Material: Clustering in anomalous files of independent particles

Ophir Flomenbom

*Flomenbom-BPS, 19 Louis Marshal St., Tel Aviv, Israel 62668*

**A – Equations of motion and results for several types of files**

In this appendix, we present the equation of motion and the main known results for a variety of files introduced in the main text: simple Brownian files, heterogeneous Brownian files, Renewal-anomalous heterogeneous files, and even Newtonain files. We also present the equation of motion of anomalous files of independent particles.

***A.1. Simple Brownian files.-*** We start with simple Brownian files. $P(x, t \mid x_0)$, the joint probability density function (PDF) for all the particles in file, obeys in a simple Brownian file, a normal diffusion equation:

$$\partial_t P(x, t \mid x_0) = D \sum_{j=-M}^{M} \partial_{x_j} \partial_{x_j} P(x, t \mid x_0). \tag{A1}$$

In $P(x, t \mid x_0)$, $x = \{x_{-M}, x_{-M+1}, \ldots, x_M\}$ is the set of particles' positions at time $t$ and $x_0$ is the set of the particles' initial positions at the initial time $t_0$ (set to zero). Equation (A1) is solved with the appropriate boundary conditions, which reflect the hard-sphere nature of the file:

$$(D\partial_{x_j} P(x, t \mid x_0))_{x_j = x_{j+1}} = (D\partial_{x_{j+1}} P(x, t \mid x_0))_{x_{j+1} = x_j} \quad ; \quad j = -M, \ldots, M-1,$$

and with the appropriate initial condition:

$$P(x, t \to 0 \mid x_0) = \prod_{j=-M}^{M} \delta(x_j - x_{0,j}). \tag{A2}$$

In a simple file, the initial density is fixed, namely, $x_{0,j} = j\Delta$, where $\Delta$ is a parameter that represents a microscopic length. The PDFs' coordinates must obey the order: $x_{-M} \leq x_{-M+1} \leq \cdots \leq x_M$. The solution of Eqs. (A1-A2) is a complete set of permutations of all initial coordinates appearing in the Gaussians [1-3],

$$P(\mathbf{x}, t \mid \mathbf{x_0}) = \frac{1}{c_N} \sum_p e^{\frac{-1}{4Dt} \sum_{j=-M}^{M}(x_j - x_{0,j}(p))^2}. \tag{A3}$$

Here, the index $p$ goes on all the permutations of the initial coordinates, and contains $N!$ permutations. From Eq. (A3), we calculated in ref. [1] the PDF of a tagged particle in the file, $P(r, t \mid r_0)$,

$$P(r, t \mid r_0) \sim \frac{1}{c_N} e^{\frac{-R_d^2}{\sqrt{2\tau}}}, \tag{A4}$$

In Eq. (A4), $R_d = r_d/\Delta$, $r_d = r - r_0$ ($r_0$ is the initial condition of the tagged particle), and $\tau = \Delta^{-2}Dt$. The MSD for the tagged particle is obtained directly from (A4):

$$<R_d^2> \sim \sqrt{2\tau}. \tag{A5}$$

Indeed, eqs. (A4)-(A5) are known results in file dynamics. In ref. (1), we suggested a new way of solving simple files, yet, interestingly, this can also solve more complicated file's types.

**A.2. Heterogeneous-Brownian files.-** The *N*-particle PDF in an heterogeneous file obeys the following equation of motion:

$$\partial_t P(\mathbf{x}, t \mid \mathbf{x_0}) = \sum_{j=-M}^{M} D_j \partial_{x_j} \partial_{x_j} P(\mathbf{x}, t \mid \mathbf{x_0}), \tag{A6}$$

with the boundary conditions:

$$(D_j \partial_{x_j} P(\mathbf{x}, t \mid \mathbf{x_0}))_{x_j = x_{j+1}} = (D_{j+1} \partial_{x_{j+1}} P(\mathbf{x}, t \mid \mathbf{x_0}))_{x_{j+1} = x_j} \quad ; \quad j = -M, \dots, M-1, \tag{A7}$$

and with the initial condition, Eq. (A2), where the particles' initial positions obey:

$$x_{0,j} = \text{sign}(j)\Delta |j|^{1/(1-a)} \quad ; \quad 0 \leq a \leq 1. \tag{A8}$$

The file diffusion coefficients are taken independently from the PDF,

$$W(D) = \frac{1-\gamma}{\Lambda}\left(\frac{D}{\Lambda}\right)^{-\gamma}, \quad 0 \leq \gamma < 1, \tag{A9}$$

where $\Lambda$ has a finite value that represents the fastest diffusion coefficient in the file. We *approximate* the solution of Eqs. (A6)-(A9) with the expression [2],

$$P(\pmb{x},t\mid \pmb{x_0}) \approx \frac{1}{c_N}\Sigma_p\, e^{-\Sigma_{j=-M}^{M}\frac{(x_j-x_{0,j}(p))^2}{4tD_j}}. \tag{A10}$$

Starting from (A10), we have calculated the PDF of the tagged particle in the heterogeneous file [2],

$$P(r,t\mid r_0) \sim \frac{1}{c_N} e^{\frac{-R_d^2}{4\tau}\tau^{\frac{(1-a)}{2-\gamma(1+a)}}} \quad ; \quad \tau = \Delta^{-2}\Lambda t. \tag{A11}$$

The MSD of a tagged particle in a heterogeneous file is taken from Eq. (A11):

$$\langle R_d^2\rangle = 2\tau^{\frac{1-\gamma}{2c-\gamma}}, \qquad c = 1/(1+a)\,. \tag{A12}$$

**A.3. Renewal-anomalous-heterogeneous files.-** In renewal-anomalous files, a random period is taken independently from a WT-PDF of the form: $\psi_\alpha(t)\sim k(kt)^{-1-\alpha}$, $0<\alpha<1$, where $k$ is a parameter. Then, all the particles in the file stand still for this random period, where afterwards, all the particles attempt jumping in accordance with the rules of the file. This procedure is carried on over and over again. The equation of motion for the particles' PDF, $P(\pmb{x},t\mid \pmb{x_0})$, in a renewal-anomalous file is obtained when convoluting the equation of motion for a Brownian file with a kernel $k_\alpha(t)$:

$$\partial_t P(\pmb{x},t\mid \pmb{x_0}) = \sum_{j=-M}^{M} D_j \partial_{x_j}\partial_{x_j}\int_0^t k_\alpha(t-u)\,P(\pmb{x},u\mid \pmb{x_0})du. \tag{A13}$$

Here, the kernel $k_\alpha(t)$ and the WT-PDF $\psi_\alpha(t)$ are related in Laplace space, $\bar{k}_\alpha(s) = \frac{s\bar{\psi}_\alpha(s)}{1-\bar{\psi}_\alpha(s)}$. (The Laplace transform of a function $f(t)$ reads, $\bar{f}(s) = \int_0^\infty f(t)\,e^{-st}dt$.) The reflecting boundary conditions accompanied Eq. (A13) are obtained when convoluting the boundary conditions of a Brownian file with the kernel $k_\alpha(t)$, where here and in a Brownian file the initial conditions are identical. We have showed in Ref. [3] that the results of renewal-anomalous files are simply derived from the results of Brownian files. Firstly, the PDF in Eq. (A13) is written in terms of the PDF that solves the un-convoluted equation, that is, the Brownian file equation; this relation is made in Laplace space:

$$\bar{P}(x, s \mid x_0) = \frac{1}{\bar{k}_\alpha(s)} \bar{P}_{nrml}(x, s/\bar{k}_\alpha(s) \mid x_0). \qquad (A14)$$

(The subscript $nrml$ stands for normal dynamics.) From Eq. (A14), it is straightforward relating the MSD of Brownian heterogeneous files and renewal-anomalous heterogeneous files [3],

$$<\bar{r}^2(s)> = \frac{1}{\bar{k}_\alpha(s)} <\bar{r}^2(s/\bar{k}_\alpha(s))>_{nrml}. \qquad (A15)$$

From Eq. (A15), it is simple showing that one can use the MSD of a file with normal dynamics in the power of $\alpha$ for obtaining the results of the corresponding renewal-anomalous file [3],

$$<r^2(t)> \sim <r^2(t)>_{nrml}^\alpha. \qquad (A16)$$

***A.4. Anomalous files of independent particles.-*** When each particle in the anomalous file is assigned with its own jumping time drawn form $\psi_\alpha(t)$ ($\psi_\alpha(t)$ is the same for all the particles), the anomalous file is not a renewal file. The basic dynamical cycle in such a file consists of the following steps: a particle with the fastest waiting time in the file, say, $t_i$ for particle $i$, attempts a jump. Then, the waiting times for all the other particles are adjusted: we subtract $t_i$ from each of them. Finally, a new waiting time is drawn for particle $i$. The most crucial difference among renewal anomalous files and anomalous files that are not renewal is that when each particle has its own clock, the particles are in fact connected also in the time domain, and the outcome is further slowness in the system (proved in the main text). The equation of motion for the PDF $P(x, t \mid x_0)$ in anomalous files of independent particles reads:

$$\partial_{t_i} P(x, t \mid x_0) = D_i \partial_{x_i} \partial_{x_i} \int_0^{t_i} k_\alpha(t_i - u_i) P(x, t'^{(i)}, u_i \mid x_0) du_i \quad ; \quad -M \leq i \leq M. \qquad (A17)$$

Note that the time argument in the PDF $P(x, t \mid x_0)$ is a vector of times: $t = \{t_i\}_{i=-M}^{M}$, and $t'^{(i)} = \{t_c\}_{c=-M, c \neq i}^{M}$. Adding all the coordinates and performing the integration in the order of faster times first (the order is determined randomly from a uniform distribution in the space of configurations) gives the full equation of motion in anomalous files of independent particles (averaging of the equation over all configurations is therefore further required). Indeed, even Eq. (A17) is very complicated, and averaging further complicates things. This is the reason that we have derived scaling-laws and applied numerical simulations for solving anomalous files of independent particles in the main text. We recall that the main result in that such files form clusters of particles for anomalous $\alpha$ that pretty much stay in the spot. This phenomenon of a phase transition depending on $\alpha$ is unique, namely, it doesn't occur in other types of files; important mentioning that the clustering can explain rafts in biological membranes and can be used in regulating biological channels, among other things. This was presented in the main text.

***A.5. Scaling laws for renewal files.-*** We present in this appendix a general formula for the MSD in any renewal file, derived from scaling law analysis in [1].

Basically, in files of hard spheres in 1D, we expect a slower propagation rate for a tagged particle in a file relative with a free particle, since a particle in a file can diffuse a net distance $l$ only when the file's particles (in the relevant direction) 'cooperate', and move in the direction of the propagation. Namely, the tagged particle's evolution is a result of a correlative motion of the system's particles. We have showed that in a file with a constant density, the scaling law for the mean absolute displacement (MAD) of the tagged particle ($<|r|>$) is written in terms of the MAD of a free particle ($<|r|>_{\text{free}}$) [1],

$$<|r|> \sim \frac{1}{n} <|r|>_{\text{free}}. \tag{A18}$$

where $n$ is the number of particles in the covered length, $<|r|>$. This equation gives the following general relation [1]:

$$<|r|> \sim \sqrt{<|r|>_{free}}, \qquad (A19)$$

since the density is fixed, $n \sim <|r|>$. Equation (A19) coincides with Eq. (A5) and Eq. (A16) for the particular cases of simple Brownian files and simple renewal anomalous files, respectively. In the particular example of Newtonian files, where $<|r|>_{free} \sim t$, we immediately find, $<|r|> \sim \sqrt{t}$, or for the MSD,

$$MSD \sim t. \qquad (A20)$$

## B – Numerical results for a large file and a file with finite size particles

In this appendix, we show results for a file with several thousand particles and for a file with finite size particles. The behavior of these files coincide with the behavior of the file reported in the main text; namely, clustering is indeed a stable phenomenon in anomalous files of independent particles and holds in small and large files and in files of point particles and of finite size particles.

*Files with finite size particles.-* Firstly, we present results in an anomalous file of independent particles of finite size. In the simulations, each particle is of a size of 0.01. (We use quantities without dimension all over.) Initially, $N$ particles are located in a symmetric way around the origin at a distance of 1.11 from each other, where $N = 2M + 1$ and here $M = 237$. Each actual jump obeys, $\sqrt{2Ddt}(2q - 1)$, where $D = 1$, $dt = 0.13$, and $q$ is a random number taken from the unit interval. After each jump the particles are ordered, and are moved such that they are not overlapping. When adjacent particles are at a distance not larger than 0.1329, they are considered clustered. We report on results for $p_n(t, \alpha)$ and $\xi(\alpha)$, $p_c(t, \alpha)$ and $\pi(\alpha)$ and $S_c(t, \alpha)$ and $\chi(\alpha)$. We recall:

- $p_n(t, \alpha)$: the number of particles in clusters over the total number of particles as a function of t and $\alpha$.
- $\xi(\alpha) \equiv p_n(t \to \infty, \alpha)$ (calculated as the average of the last 10% of the trajectory)
- $p_c(t, \alpha)$: the number of clusters over the total number of particles as a function of t and $\alpha$
- $\pi(\alpha) \equiv p_c(t \to \infty, \alpha)$ (calculated as the average of the last 10% of the trajectory)
- $S_c(t, \alpha)$: the average number of particles in a cluster as a function of t and $\alpha$
- $\chi(\alpha) \equiv S_c(t \to \infty, \alpha)$ (calculated as the average of the last 10% of the trajectory)

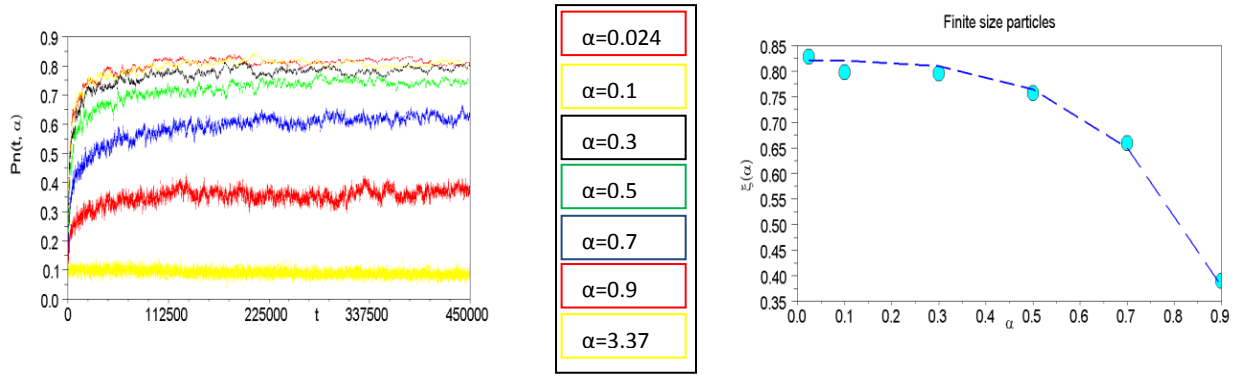

**FIG B1** $p_n(t, \alpha)$ (left panel) as a function of the cycle index t for 7 values of $\alpha$ (specified in the middle panel) and its steady state function $\xi(\alpha)$ (right panel) from trajectories of files of anomalous, independent, particles of finite size. In the simulations, $N = 501$, the size per particle is 0.01, $\Delta = 1.11$ (the initial distance among particles), $D = 1$ (the diffusion coefficient of the particles), $dt = 0.13$, and the jumping distance obeys, $\sqrt{2Ddt}(2q - 1)$, where $q$ is a random number uniform in the unit interval. (Here, we use unite without dimensions.) Reflecting boundaries are positioned at a distance of 27 integers relative to the initial position of the particles at the edges, in the direction that extends the interval. When adjacent particles are at a distance not larger than 0.1329, they are considered clustered. $p_n(t, \alpha)$ and $\xi(\alpha)$ have the same form seen in the case of point particles, yet here, $\xi(\alpha \to 0)$ has a smaller value of about 5 percent relative to a file of point particles.

We note that the results reported here were tested with several types of algorithms for simulating finite size particles. It is important using a continuous coordinate and a continuous-motion-technique (jumps are made and then the particles are ordered and moved such that they are not overlapping). Nevertheless, this is the most physical way of simulating finite size particles in a file, since particles in nature move in a continuous way, and when they bump each other they can exchange momentum for an incremental distance (in solution of finite temperature). Figure B1 shows $p_n(t, \alpha)$ (left panel) as a function of t for several values of $\alpha$, $\alpha = 0.024, 0.1, 0.3, 0.5, 0.7, 0.9, 3.37$, and its steady function $\xi(\alpha)$ as a function of $\alpha$ (right panel). The forms of these quantities coincide with the forms of the corresponding quantities in files of point particles. In particular, the fitting function for $\xi(\alpha)$, $\tilde{\xi}(\alpha)$, follows the form,

$$\tilde{\xi}(\alpha) = 0.8492 \left(1 - \left(\frac{\alpha}{0.99}\right)^{3.09}\right)^{0.537} - 0.028. \tag{B1}$$

This is a particular case of the general formula used also for point particles,

$$\tilde{\xi}(\alpha) = c_1 \left(1 - \left(\frac{\alpha}{c_2}\right)^{\mu_{\tilde{\xi}}}\right)^{\nu_{\tilde{\xi}}} - c_3. \tag{B2}$$

Excluding the first parameter $c_1$, that here equals, $c_1 = 0.8492$ and equals, $c_1 = 0.98$, for point-like particles, $\tilde{\xi}(\alpha)$ is identical in both cases.

Now, figure B2 shows the behavior of $p_c(t, \alpha)$ and $\pi(\alpha)$, and $S_c(t, \alpha)$ and $\chi(\alpha)$ for finite size particles. The fitting function for a file of $\pi(\alpha)$, $\tilde{\pi}(\alpha)$, follows the form,

$$\tilde{\pi}(\alpha) = 0.82 \left(1 + \left(\frac{\alpha}{1.19}\right)^{2.49}\right)^{0.29} - 0.75. \tag{B3}$$

Again, this fitting function $\tilde{\pi}(\alpha)$ obeys the exact same general form as in a file of point-like particles,

$$\tilde{\pi}(\alpha) = p_1\left(1 + \left(\frac{\alpha}{p_2}\right)^{\mu_{\tilde{\pi}}}\right)^{\nu_{\tilde{\pi}}} - p_3. \tag{B4}$$

Here the difference is in $p_1$ ($p_1 = 0.82$ here and $p_1 = 0.78$ in a file of point-like particles) and in $\nu_{\tilde{\pi}}$ ($\nu_{\tilde{\pi}} = 0.29$ here and $\nu_{\tilde{\pi}} = 0.42$ in point-like particles). The fitting function for the average number particles in a cluster $\chi(\alpha)$, $\tilde{\chi}(\alpha)$, obeys,

$$\tilde{\chi}(\alpha) = (12.29 - 11\alpha)/N. \tag{B5}$$

Again, this is the a particular case of the general form,

$$\tilde{\chi}(\alpha) = (\tilde{\chi}_0 - x\alpha)/N, \tag{B6}$$

found also for point particles. Both parameters of the linear function $\tilde{\chi}(\alpha)$ are three times smaller here compared with the case of point particles; mathematically, this is a direct consequence of the difference in the parameters of $\tilde{\pi}(\alpha)$ among the files: since the number of clusters is larger here (2.9 times larger), yet the total number of particles in clusters is about equal, the average number of particles in a cluster is smaller (about one third).

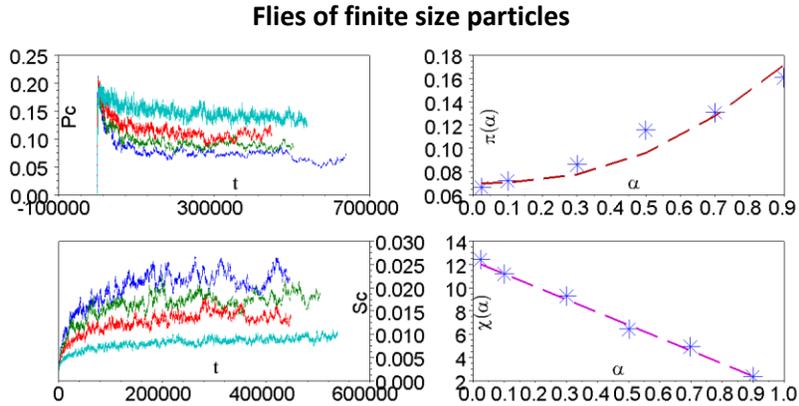

**FIG B2** *Finite size particles.-* $p_c(t, \alpha)$ (upper left panel) as a function of the cycle index t for several values of $\alpha$ ($\alpha = 0.024, 0.3, 0.5, 0.7$ from the lower curve) and its steady state function $\pi(\alpha)$ (upper right panel), $S_c(t, \alpha)$ (lower left panel, where here, curves with smaller values of $\alpha$ are on top) and its steady state function, $\chi(\alpha)$. $\pi(\alpha \to 0)$ is about three times larger in finite size particle file, and $\chi(\alpha)$ is three times smaller.

*A larger file.-* Here, we examine the behavior of a file with several thousand particles. Initially, $N$ particles are located in a symmetric way around the origin at a distance of 1 from each other, where $N = 2M + 1$ and here $M = 1137$ (this file is about five times larger than the file presented in the main text). The fastest particle attempts a jump, and the random times are adjusted. Each actual jump obeys, $\sqrt{2Ddt}(2q - 1)$, where $D = 1$, $dt = 0.13$, and $q$ is a random number taken from the unit interval. After each jump the particles are ordered. Reflecting boundaries were placed at a distance of unity from the particles at the edges in the direction that extends the interval. When adjacent particles are at a distance not larger than 0.1, they are considered clustered. Results were collected for the following values of $\alpha$,

$$\alpha = 0.1, 0.3, 0.5, 0.7, 0.9, 3.37. \tag{B7}$$

The results are reported in Fig. B3 showing $\xi(\alpha)$, $\pi(\alpha)$, and $\chi(\alpha)$.

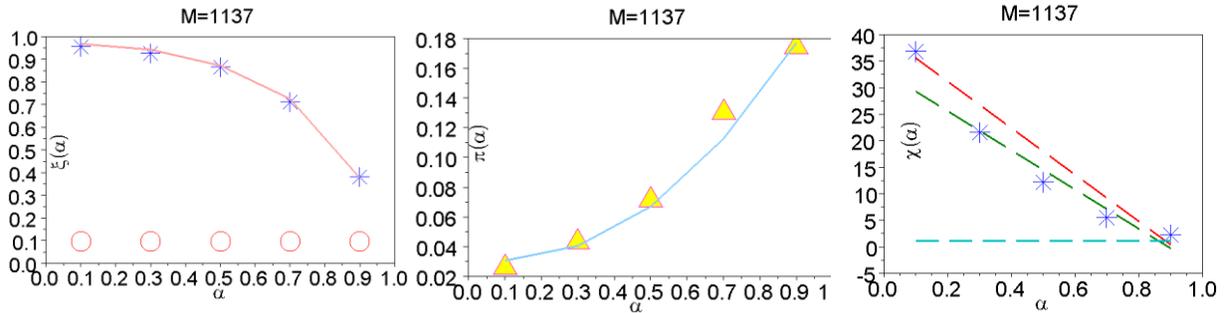

**Figure B3** $\xi(\alpha)$, $\pi(\alpha)$, and $\chi(\alpha)$ for a file of 2274 point particles. The left panel shows $\xi(\alpha)$ (blue asterisk) the fitting function $\tilde{\xi}(\alpha)$ in equation (B8) and the level of clustering in a file with $\alpha = 3.37$. The middle panel shows $\pi(\alpha)$ (yellow triangles) and the fitting function $\tilde{\pi}(\alpha)$ presented in Eq. (B9). The clustering level in a file with $\alpha = 3.37$ is about 0.1, and is not shown as it represents only very small clusters and doesn't reflect the clustering phenomenon that the function $\pi(\alpha)$ represents (small values of $\pi(\alpha)$ stands for very large clusters). The right panel shows $\chi(\alpha)$ (blue asterisks) the fitting function in Eq. (B10), red dashed curve, the fitting function $\tilde{\chi}(\alpha)$ of the smaller file (green dashed curve), and the average size of a cluster in a file with $\alpha = 3.37$ (the light blue dashed line).

The fitting functions of these quantities obey the same general formula introduced in Eqs. (B2), (B4), B(6). In particular, we find the following fitting functions:

- $\tilde{\xi}(\alpha) = 1.1 \left(1 - \left(\frac{\alpha}{0.96}\right)^{2.416}\right)^{0.4} - 0.13.$ (B8)

- $\tilde{\pi}(\alpha) = 0.98 \left(1 + \left(\frac{\alpha}{1.15}\right)^{2.2}\right)^{0.3} - 0.95.$ (B9)

- $\tilde{\chi}(\alpha) = 39.99 - 43.99\alpha.$ (B10)

These results are in accordance with the results of the smaller file.

### Appendix C – anomalous files in two dimensions

Here, we test the occurrence of the clustering phenomenon in an ordered network of files embedded in two dimensions forming an isotropic network of files and junctions. This system can mimic diffusion in a membrane that has many obstacles for the diffusing objects, forming streets and junctions. We show that clustering occurs also in this system. Indeed, further study of the dynamics of independent anomalous particles in files embedded in two-dimensions and even in three-dimensions is needed; still, from the results reported here we can related the critical phenomenon of clustering with rafts in membranes. This was discussed in the main text, where in this appendix we present the results from the simulations.

In files embedded in two dimensions, we have two free parameters defining the system: the percentage of intersections and the length of the interval until an intersection occurs. In the current system, files intersect each other only 1% of the time, every interval of 10 (see figure C1 for an illustration). Indeed, the results are sensitive to the branching parameters: when branching occurs in smaller intervals, clustering decreases, and we can speculate that when diffusion happens in two dimensions, the clustering phenomenon is not observed when the density is reasonable (not too high). This is in accordance with known results showing that the slower diffusion so typical for a particle in a file in one-dimension does not hold for diffusion of hard particles in two-dimensions, where in such a system standard diffusion is seen (when the density is not too high). Still, we have chosen here reasonable parameters for the branching: the average size of a jump is 0.25, and the branching occurs every interval of 10; this is not too small interval so branching indeed has a role (seen also in the results), yet the branching happens after frequent enough jumps and the clustering is indeed seen.

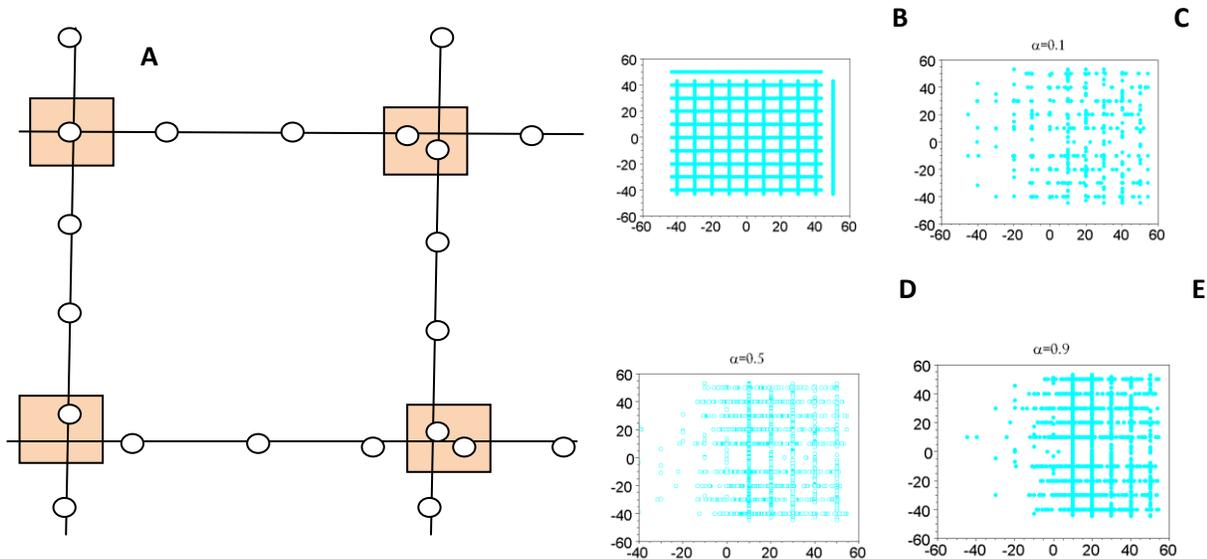

**Figure C1.** (**A**) An illustration of files embedded in two dimensions, with 1% branching every interval of 10. At the end of every interval of length 10 in a given direction there is a possibility for branching for a length of 0.1 (the colored squares). Here, particles are marked with circles. (**B-E**) Here, we present the particles at the initial stage of the process (**B**) and at the end of process (after about 733 thousands cycles), for $\alpha = 0.1$ (**C**) $\alpha = 0.5$ (**D**), and $\alpha = 0.9$ (**E**). In these plots, a cluster is seen as a dense area of particles with areas free of particles around it.

Mathematically, we define files in two dimensions in an area $A(x, y)$: in $x$-files, $x$ is a continuous coordinate, and $y$ is an integer, $y = 0, \pm 4, \pm 8, \pm 12, ...$ In $y$-files, set $x \leftrightarrow y$. In the simulations, 1980 particles were located in 20 one-dimensional files: 10 $x$-files (constant $y$) and 10 $y$-files (constant $x$). In each file there are 99 particles; the particles are located around the origin, in a symmetric way, a particle every 0.88. At about, $x, y = \pm 52$, we put reflecting boundaries. See **Fig. C1** for an illustration of the system, the initial configuration and the final configuration for various values of $\alpha$ in the area $A(x,y)$.

The quantities $p_n(t,\alpha)$ and $\xi(\alpha)$ and $p_c(t,\alpha)$ and $\pi(\alpha)$ and $S_c(t,\alpha)$ and $\chi(\alpha)$ are calculated for this system. Results are obtained for,

$$\alpha = 0.1, 0.3, 0.5, 0.7, 0.9, 3.37. \qquad (C1)$$

Firstly, **Fig. C2** shows the quantities $p_n(t,\alpha)$, $p_c(t,\alpha)$ and $S_c(t,\alpha)$. These quantities are calculated in a similar way for files embedded in one dimension: each one-dimensional file (constant $x$ or constant $y$) is calculated as an independent file, clustering is when particles are at a distance of 0.11 or smaller, and then an average is applied. The familiar forms of these quantities seen in files embedded in one dimension are observed also in files embedded in two-dimensions with 1% branching every interval of 10. Namely, clustering occurs even when the files are embedded in two dimensions.

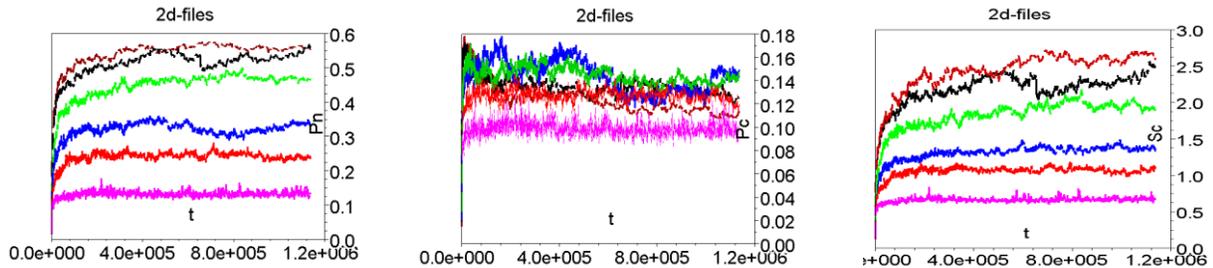

**FIG C2** $p_n(t,\alpha)$, $p_c(t,\alpha)$ and $S_c(t,\alpha)$ for anomalous files of independent particles embedded in two dimensions with 1% branching every interval of length 10. In the left panel, shown is $p_n(t,\alpha)$ for $\alpha$ values presented in Eq. (C1) (curves with larger values of $\alpha$ are of smaller height). The middle panel shows $p_c(t,\alpha)$; each curve with a particular $\alpha$ has the same color used its counterpart $p_n(t,\alpha)$ shown in the left panel. Here, the number of clusters is about the same for all values $\alpha$, and this means that the clustering is a bit weaker in this system relative with the clustering of a file embedded in one dimension. There are a lot of free particles and a lot of small clusters. This is seen explicitly in the right panel when plotting the number of particles in a cluster versus t for the various values of $\alpha$. Clusters are indeed smaller relative with files embedded in one dimension, still there is a prominent effect of an increase in the average number of particles in a cluster when $\alpha$ decreases.

Figure C3 quantifies further the clustering in files in two dimensions when showing the quantities $\xi(\alpha)$, $\pi(\alpha)$ and $\chi(\alpha)$. The fitting functions show that the branching affects the degree of clustering: here, 55% of the particles (relative to a file in one dimension) are in clusters. Still, when comparing the results in the 2d system for the various values of $\alpha$, we see that when $\alpha \to 0$, the percent of clustering is five times larger compared with $\alpha = 3.37$. This indicates on a prominent clustering in anomalous files of independent particles in 2d relative with other files' types in 2d.

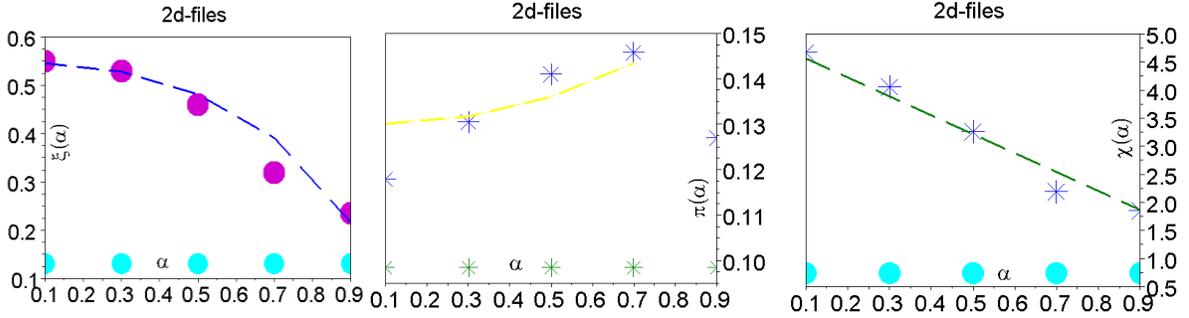

**Figure C3** $\xi(\alpha)$, $\pi(\alpha)$, and $\chi(\alpha)$ for files embedded in two dimensions with 1% branching every interval of 10. In every figure the horizontal curve stands for the result for $\alpha = 3.37$. The fitting curves for $\pi(\alpha)$ is obtained for the first 4 points excluding the fifth. For the specific forms of the fitting functions, see Eqs. (C2)-(C4). These plots show that clustering is indeed seen in 2d-files with 1% branching every interval of 10.

The fitting functions shown in Fig C3 obey the formulae in Eqs. (B2), (B4), (B6), of the file embedded in one dimension:

$$\tilde{\xi}(\alpha) = 0.55\left(1 - \left(\frac{\alpha}{0.99}\right)^{2.374}\right)^{0.577} - 0.0028. \tag{C2}$$

$\tilde{\pi}(\alpha)$ follows:

$$\tilde{\pi}(\alpha) = 0.13\left(1 + \left(\frac{\alpha}{1.19}\right)^{2.505}\right)^{0.42}. \tag{C3}$$

$\tilde{\chi}(\alpha)$ follows:

$$\tilde{\chi}(\alpha) = 4.9 - 3.37\alpha. \tag{C4}$$

REFERECNES